\documentclass[a4paper,11pt]{article}
\pdfoutput=1 
\usepackage{jheppub} 
\usepackage{graphicx,amssymb,amsmath,color}
\usepackage{hyperref}
\usepackage{multirow}
\usepackage{graphicx}
\usepackage{float}
\usepackage{amssymb}
\usepackage{xcolor,soul}

\colorlet{RED}{red}
\colorlet{OLIVE}{olive}
\setstcolor{red}

\newcommand{\beq}{\begin{equation}}
\newcommand{\eeq}{\end{equation}}
\def\bea{\begin{eqnarray}}
\def\eea{\end{eqnarray}}
\newcommand{\bei}{\begin{itemize}}
\newcommand{\eei}{\end{itemize}}
\newcommand{\bigO}{\mathcal{O}}

\newcommand{\Fig}[1]{Fig.~\ref{#1}}
\newcommand{\Eq}[1]{Eq.~(\ref{#1})}
\newcommand{\Sec}[1]{Sec.~\ref{#1}}
\newcommand{\App}[1]{Appendix~\ref{#1}}

\def\={\,=\,}
\def\+{\,+\,}
\def\-{\,-\,}
\def\Mch{{\cal M}}
\def\Msun{M_\odot}

\begin{document}

\title{
Probing Cosmic Strings with Gravitational-Wave Fringe
}

\author[a]{Sunghoon Jung,}
\author[a]{TaeHun Kim}

\affiliation[a]{Center for Theoretical Physics, Department of Physics and Astronomy, \\Seoul National University, Seoul 08826, Korea}

\emailAdd{sunghoonj@snu.ac.kr}
\emailAdd{gimthcha@snu.ac.kr}


\abstract{Cosmic strings are important remnants of early-Universe phase transitions. We show that they can be probed by Gravitational Waves (GWs) from compact binary mergers.
If such chirping GW passes by a cosmic string, it is gravitationally lensed and left with a characteristic signal of the lensing -- the GW fringe. It is observable naturally through the frequency chirping of GWs. 
This allows to probe cosmic strings with small tension $\Delta = 8\pi G \mu = 10^{-6} \text{ -- } 10^{-10}$, just below the current constraint, at high-frequency LIGO-band and mid-band detectors. Although its detection rates are estimated to be small, even a single detection can be used to identify a cosmic string.
Contrary to the stochastic GW produced from loop decays only in local $U(1)$ models, the GW fringe can directly probe straight strings model independently. This is also complementary to the existing probes with the strong lensing of light. 
}

\maketitle


\section{Introduction}

It is believed that the early Universe has evolved down to the Standard Model of particle physics from a more unified or fundamental theory by experiencing several phase transitions. The spontaneous symmetry breaking of a (global or gauge) $U(1)$ symmetry must have produced cosmic strings~\cite{Kibble:1980mv}. Cosmic strings are one-dimensional topological field configurations, produced also from string theory and vortex-like solutions of quantum field theory. Once produced, they gradually evolve into the scaling regime, where the total energy density of strings and closed string loops remains constant with the expansion of the Universe~\cite{Vanchurin:2005pa,Rocha:2007ni,Kibble:1976,Vilenkin:1984ib}. Thus, observing cosmic strings that remain today can give important clues on the physics of the early Universe.

Cosmic strings are characterized by its tension $\mu$ (energy density per unit length), given by $G \mu = v^2/m_{\rm Pl}^2$ with the symmetry-breaking vacuum expectation value $v$ of a $U(1)$ symmetry. The thickness of a gauge string is small of order $1/v$ so that it is a highly localized one-dimensional energy clump. The global string has its energy more spread in space stored in the Goldstone fields wrapping around the string, but their gravitational effects on null rays (photons and GWs as in this paper) are almost equivalent up to a marginal logarithm factor that describes the spatial spread~\cite{Harari:1988wa}. As string's high local energy density disturbs the homogeneity of the CMB power spectrum, the tension is constrained to be $G\mu \lesssim 10^{-7}$ (or $\Delta \equiv 8\pi G\mu \lesssim 10^{-6}$ in our notation)~\cite{Charnock:2016nzm}. Although cosmic strings consequently cannot be the dominant fraction of dark matter (DM) or the seed of structure formation, their high local energy density may still leave important observable signatures.

Cosmic strings have been probed mainly by CMB anisotropies, stochastic GWs radiated from gauge string loops, and by their gravitational lensing on lights. The GW radiation from string loops is an essential energy-loss mechanism in the evolution of a string network into the scaling regime, where the string network's energy density can remain safely small~\cite{Vanchurin:2005pa,Rocha:2007ni,Kibble:1976,Vilenkin:1984ib}. Thus, the stochastic GW is a prime observable of cosmic strings that is actively searched for~\cite{Kuroyanagi:2012jf,Blanco-Pillado:2017rnf,Ringeval:2017eww,Abbott:2017mem} (this may lead to a constraint as strong as $G\mu \lesssim 10^{-11}$, albeit model dependencies), and its spectrum can also probe the cosmological evolution history of the Universe~\cite{Cui:2017ufi,Cui:2018rwi}. On the other hand, global strings evolve into the scaling regime through the rapid decay of loops by radiating off Goldstone bosons, so the stochastic GW signal is absent. Instead, what always remains model independently is the order-one number of long (at least Hubble-sized) strings per horizon~\cite{Consiglio:2011qe,Harari:1987ht,Sikivie:1990sb}, as the causality requires. The long string can be directly probed by the strong lensing of light (multiple images resolved in angle or arrival time)~\cite{Vilenkin:1981zs,Gott:1984ef} and femto-lensing of gamma-ray bursts (GRBs)~\cite{Yoo:2012dn,Suyama:2005ez}, but the detection rates are small as will be discussed. Therefore, new independent probes are needed to boost the detection prospects.

In this paper, we study the chirping GW from binary mergers as a probe of cosmic strings. The new observable is the interference fringe of GWs produced by the gravitational lensing of cosmic strings; this is similar to the GW fringe produced by compact dark matter such as primordial black holes~\cite{Jung:2017flg,Nakamura:1997sw}. The fringe is naturally observed through the chirping (particular pattern of time-dependent frequency evolution) of GWs. Thus, the sensitivity range of $\Delta$ depends on both the frequency band and frequency resolution of measurements.
We will show that the LIGO frequency band ($5 \text{ -- } 5000$ Hz) and the mid-frequency band ($0.1 \text{ -- } 10$ Hz) are particularly useful, as they can probe $\Delta \lesssim 10^{-6}$ just below the current constraint. 

We start by calculating the lensing fringe and its detection rate in \Sec{section lensing fringe} and \Sec{section calculation}, respectively. Then we present numerical results and analyze detection prospects in \Sec{section prospects}, discuss further phenomenology in \Sec{section discussion}, and conclude at the end.

\section{Lensing fringe from cosmic strings} \label{section lensing fringe}

\subsection{Straight strings} \label{section straight strings}

The space-time geometry around a straight (gauge) string is described by a conical space~\cite{Vilenkin:1981zs}  
\beq
ds^2 \= dt^2 \- dZ^2 \- dR^2 \- \left(1- \frac{\Delta}{2\pi} \right)^2 R^2 d \hat{\phi}^2,
\eeq
where the string is placed along the $Z$-axis and $\hat{\phi}$ is measured around the string. By the redefinition of the azimuthal angle $\phi = (1-\Delta/2\pi) \hat{\phi}$, the conical space can be viewed as the Euclidean flat space with a deficit angle $\Delta \equiv 8 \pi G \mu$. The deficit angle is defined by a boundary condition on the GW amplitude $h(\phi= 0)=h(\phi= 2\pi-\Delta)$ in the plane perpendicular to the string, with the allowed range $0 \leq \phi \leq 2\pi - \Delta$. The gravitational effect of global strings on null rays is described by a similar deficit angle~\cite{Harari:1988wa}; thus, we simply use the same metric and lensing calculation for both cases.

There is a freedom to choose the direction of $\phi=0$ in mapping the conical space to the Euclidean space with deficit angle. In the presence of a GW source, it is particularly convenient to choose such that $\phi=0$ is mapped to point to the source in the conical space. In the Euclidean space, this is equivalent to the source located at $\phi=0$ and $\phi= 2\pi - \Delta$ simultaneously, which guarantees the boundary condition to be satisfied. Now, within the allowed range of $\phi$, null rays from the source are propagated according to the usual Helmholtz equation on the Euclidean space. Consequently, the GW rays arriving at the observer can be obtained by the Kirchhoff diffraction integral of freely propagating rays around the string \cite{schneider:1992}. 

\medskip

The gravitationally lensed GW waveform (that an observer measures) is parameterized in the frequency domain as 
\beq
\widetilde{h}^L(f) \= \widetilde{h}(f)\, F(f)\, e^{i f \phi_m}. \label{equation htilde lensed}
\eeq
Here, $\widetilde{h}(f)$ is the unlensed waveform which is explained in \Sec{sec:calculation}. The complex lensing amplification $F(f)$ is the solution of the diffraction integral \cite{Yoo:2012dn} 
\begin{eqnarray}
F(f) &=& e^{-i\frac{fw}{2}(1+2y)} \left\{1-\frac{1}{2} \text{erfc}\left[\sqrt{\frac{fw}{2i}} (1+y)\right] \right\} \nonumber \\ 
&& \+ e^{-i\frac{fw}{2}(1-2y)} \left\{1-\frac{1}{2} \text{erfc}\left[\sqrt{\frac{fw}{2i}} (1-y)\right] \right\}, \label{equation F}
\end{eqnarray}
where 
\begin{equation}
    w \, \equiv \, 2\pi  \frac{d_L d_{LS}}{d_S} \left(\frac{\Delta}{2}\right)^2 
    \, \simeq \, 1.6 \, {\rm sec} \, \times \left(\frac{d_L}{100 \ \text{Mpc}}\right)\left(\frac{d_{LS}}{100 \ \text{Mpc}}\right)\left(\frac{100 \ \text{Mpc}}{d_S}\right)\left(\frac{\Delta}{10^{-8}}\right)^2  \label{equation w}
\end{equation}
is the characteristic path difference among the GW rays. $y \equiv 2 \phi_{S} / \Delta$, where $\phi_S$ is the azimuthal angle $\phi$ between the two lines, one connecting the observer and the string and the other connecting the string and the source, on the plane perpendicular to the string. As physics is symmetric with respect to $y\rightarrow -y$, hereafter we deal with only $y>0$. We denote the angular diameter distance by $d$, luminosity distance by $D$, and comoving distance by $\chi$; the subscripts $L,S$ and $LS$ denote the distances to the lens, source and between the lens and source, respectively. Since distances and $\Delta$ are combined into one variable $w$ on which lensing amplification $F(f)$ depends, they cannot be measured separately from the observed waveform alone. Lastly, the extra phase in \Eq{equation htilde lensed}
\beq
\phi_m \= \frac{w}{2} + wy  \label{eq:tc}
\eeq
makes the arrival time of the fastest path to zero so that we can focus only on the relative phases among the rays; this shall be consistent with our choice of $t_c=0$ in the GW waveform as discussed in \Sec{sec:calculation}. 

\begin{figure}[t]
    \centering
    \includegraphics[width=0.65\textwidth]{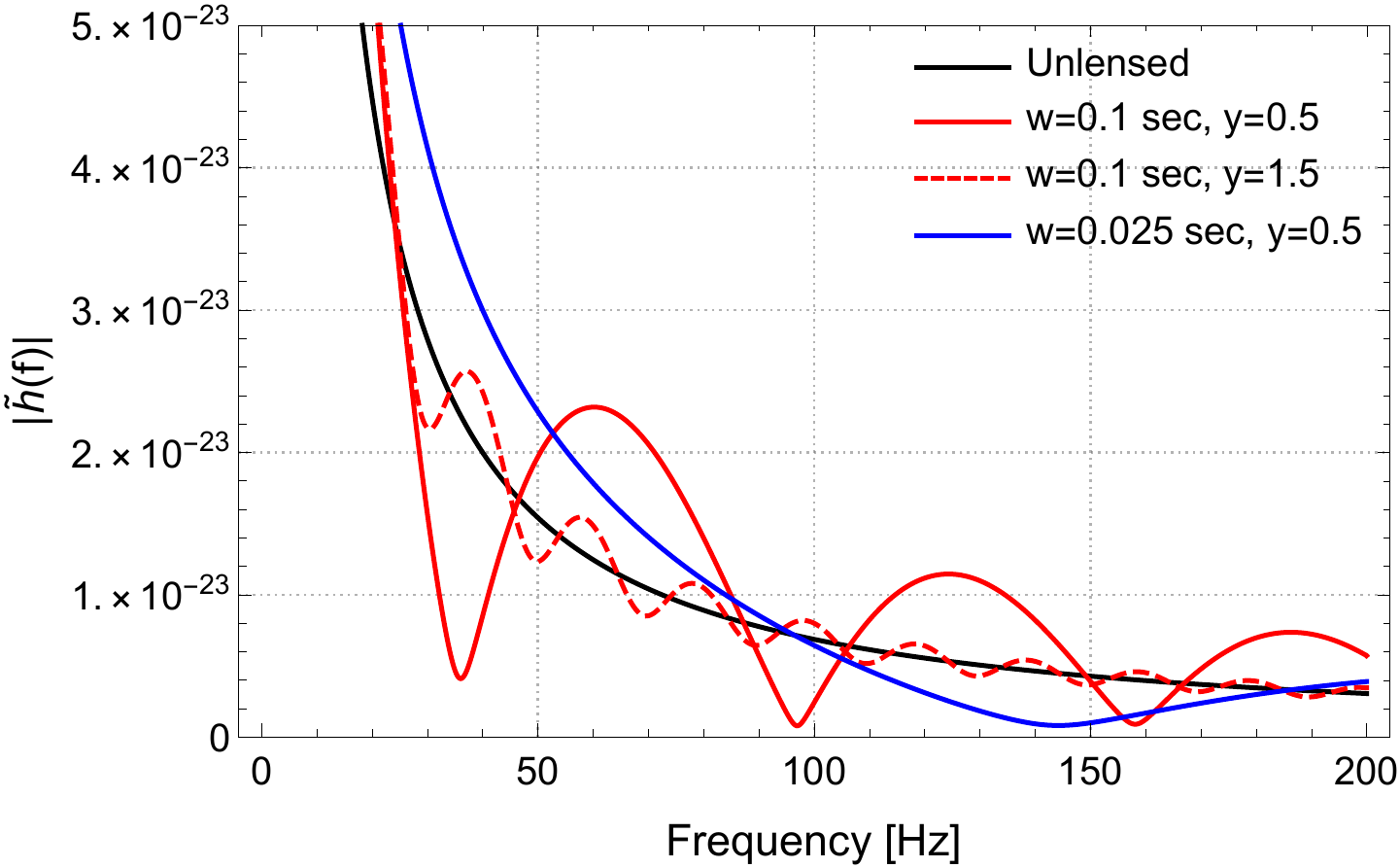}
    \caption{GW fringes from cosmic-string lensing in the frequency domain. The benchmark GW is from 30$M_\odot$-30$M_\odot$ binary at 400 Mpc of luminosity distance: unlensed (black), lensed by a cosmic string with $w=0.1$ sec and $y=0.5$ (red-solid), $w=0.1$ sec and $y=1.5$ (red-dashed),  $w=0.025$ sec and $y=0.5$ (blue). The $w$ and $y$ are the characteristic path difference and angular impact parameter, respectively, defined in \Eq{equation w} and below.
    }
    \label{figure waveform}
\end{figure}

\medskip
In \Fig{figure waveform}, we show an example unlensed waveform $|\widetilde{h}(f)|$ and its lensed waveforms $|\widetilde{h}^L(f)|$ in various lensing environments. The lensed waveforms oscillate with respect to the unlensed waveform, having local maxima at regular frequency intervals. The oscillation is due to alternating constructive and destructive interferences between the rays contributing to the $F(f)$. We name this \emph{interference fringe} as the ``GW fringe''~\cite{Suyama:2005ez, Nakamura:1999ptps,Jung:2017flg}. It has characteristic features that not only allow efficient detection but also distinction from other chirping effects and other kinds of lenses. 

\Fig{figure waveform} shows those characteristic features of GW fringes from cosmic strings, in their oscillation amplitudes and widths in the frequency domain, with three example cases. For $y<1$, the interference occurs maximally so that $|F|$ oscillates from nearly 0 to nearly 2 in the whole range of $f$, for any values of $w$ (red and blue solid lines with $y=0.5$ in \Fig{figure waveform}). On the other hand, for $y>1$, the amplitude of the interference becomes smaller for larger $f$, $w$, and $y$. The amplitude decreasing with $f$ shown in the red-dashed ($y=1.5$) is a character of the diffraction.

These can be analytically understood in the limit $f w \gg 1$ (or, more precisely, $f w(1 \pm y)^2 \gg 1$), in which the phase shifts among the rays ($\sim f w$) span many numbers of the fringe period. In this limit, the amplification factor becomes the interference among \emph{three} rays:
\beq
F(f) \, \simeq \,  e^{-i\frac{fw}{2}(1+2y)} \+ \frac{1+\text{sgn}(1-y)}{2} e^{-i\frac{fw}{2}(1-2y)} \- \frac{2}{\sqrt{2\pi fw} (1-y^2)} e^{i (\frac{fw}{2} y^2 +\frac{\pi}{4})}.\label{equation F approx}
\eeq
The first two terms are identified as the result in the geometrical optics limit and the last term is a diffracted ray. For $y<1$, the two geometrical-optics terms are deflected rays in each side of the string; equivalently, in the Euclidean space with a deficit angle, the two rays arrive at us straightly from the two images of the source at $\phi=0$ and $\phi=2\pi - \Delta$. They correspond to the usual two rays in the geometrical optics limit of the point-mass lensing, but one of them (the second term) has its amplitude decreased by half for $y=1$ and disappears for $y>1$.

The third term in \Eq{equation F approx} is the ray diffracted from the cosmic string. The diffraction nature is encoded in both the amplitude and the phase. The phase (or path difference) implies that this ray first hits the string and then propagates to the observer (actually spreads out all directions). The amplitude decreasing with $f$, $\propto 1/\sqrt{fw}(1-y^2)$, is indeed a common feature of diffraction of any waves; it is remarked that this term is exactly derived by the geometric theory of diffraction \cite{Keller:1962josa}. Such $f$ dependence of the amplitude for $y>1$ is shown in \Fig{figure waveform}.

After all, all three rays interfere\footnote{Readers may refer to Fig. 3 of Ref.~\cite{Fernandez-Nunez:2016cea} for the three rays projected onto the plane perpendicular to the string (with cautions for slightly different notations).}. For $y<1$, all three rays reach the observer, and the two geometric rays make full interference while the effect of a diffracted ray is relatively suppressed. For $y>1$, although the diffracted ray weakens with increasing $y$, this produces the GW fringe. So in this case, the amplitude of a fringe is proportional to the amplitude of a diffracted ray, $1/\sqrt{fw}(1-y^2)$.

\medskip

Another more important feature in \Fig{figure waveform} is that all three cases have constant fringe widths in the frequency domain. The width $f_{\rm width}$ is determined by phase shifts among the major interfering rays -- two geometric rays for $y\leq1$ and one geometric ray and the diffracted ray for $y>1$ -- such that $f_{\rm width}$ times the path difference equals to unity. By reading the phases in \Eq{equation F approx}, we obtain
\begin{eqnarray}
    f_{\rm width} \, \simeq \,
    \begin{cases}
        \pi/wy,& \text{for } y\leq 1\\
        4\pi/w(1+y)^2,& \text{for } y>1
    \end{cases}. \label{equation fringe width analytic}
\end{eqnarray}
For example, the width $f_{\rm width} \simeq 4$ Hz for $\Delta=10^{-8}$, $d_L = d_{LS} = 100$ Mpc, $d_S=200$ Mpc, and $y=1$ (giving $w \simeq 0.8$ sec). The width is indeed constant in $f$ for the given lensing situation, so that fringes repeat with a constant period in the frequency domain as discussed and shown in \Fig{figure waveform}. This is a general feature of lensing fringes~\cite{Jung:2017flg} as the characteristic time-delay is determined by the geometry. But cosmic-string fringes also have features distinct from point-mass fringes as will be discussed in \Sec{section distinction}. The constant width is also a key property that allows efficient discrimination against various oscillating effects of chirping as will be discussed in \Sec{sec:calculation}.

\subsection{Loops} \label{section loops}
The gauge string is accompanied by loops, which can also give rise to lensing effects. In this subsection, we first calculate the total length of the loops and then show that its large fraction can be treated as straight strings in the lensing perspective. 

\medskip

The string-loop network in the scaling regime is assumed to be the one-scale model \cite{Caldwell:1991jj} with the matter-dominated universe. In this model, loops have initial length $L(t_B) = \alpha l(t_B)$ at their birth time $t_B$, where $l(t)$ is the particle horizon at time $t$ and $\alpha$ is a free parameter with a wide range $\alpha = 0.1$ -- $10^{-5}$~\cite{Caldwell:1991jj} (we choose a specific value for numerical results at the end of this section). Once a loop is formed, it radiates GW and shrinks with the energy-loss rate $dE/dt = -\Gamma G \mu^2$, where $\Gamma \approx 50$ is a constant determined by numerical simulation \cite{Caldwell:1991jj}. Then the distribution of length-weighted comoving number density at some given time $t$ is~\cite{Mack:2007ae}  
\begin{equation}
    L\frac{dn(t)}{dL} \,\approx\, C' H_0^2 \frac{L}{\left(L+\frac{1}{3}\Gamma G \mu l(t)\right)^2}, \label{equation loop distribution}
\end{equation}
where $n(t)$ is the total comoving number density of the loops, $H_0$ is the Hubble parameter today, and $C'$ is a constant to be determined by numerical simulations. In this derivation, $l(t) = 3t$ for the matter dominated universe and $\alpha \gg \Gamma G \mu$ are used. The time dependence kept until this step finally drops out in the energy density fraction $\Omega_{\rm loop}$, as should be in the scaling regime
\begin{equation}
    \Omega_{\rm loop} \,\approx\, \frac{8\pi}{3}G\mu C'\left[\ln\left(\frac{3\alpha}{\Gamma G \mu}\right)-1\right].
\end{equation}
Meanwhile, the number of straight strings (at least as long as the Hubble length) in a Hubble volume was also calculated by simulations \cite{Caldwell:1991jj, Consiglio:2011qe}, yielding ${\cal O}(1)$ albeit some uncertainties; similarly, the number of global strings is also estimated to be ${\cal O}(1)$ per Hubble volume~\cite{Harari:1987ht,Sikivie:1990sb}. In this work, we simply take the string number density to be one in a Hubble volume, so that $\Omega_{\rm cs} = 8\pi G\mu/3$ (subscript cs refers only to the straight strings). Therefore, the energy ratio of total loops to total straight strings is
\begin{equation}
    \frac{\Omega_{\rm loop}}{\Omega_{\rm cs}} \= C'\left[\ln\left(\frac{3\alpha}{\Gamma G \mu}\right)-1\right]. \label{eq:OmegaRatio}
\end{equation}
This is rather a generic result of the one-scale model. We use $\Gamma = 50$ and $C' = 0.625$, where the latter choice is the one that we found to fit the recent simulation result~\cite{Blanco-Pillado:2013qja}. 
The ratio is minimal $\Omega_{\rm loop}/\Omega_{\rm cs} \approx 1.1$ for $\Delta = 10^{-6}$ and $\alpha = 10^{-5}$ and grows to $\approx 12.6$ for $\Delta = 10^{-10}$ and $\alpha = 0.1$. Thus, for the most part of the parameter space, a large fraction of string network's energy reside in the loop.

\medskip

What kind of lensing effects do loops produce? Although a loop at a far distance produces the Schwarzschild metric of the total loop mass, it can be treated as a straight string in its vicinity. Roughly speaking, the loop produces the same lensing effects as straight strings when the loop segment produces a pair of double images of a source just behind it. A loop at angular-diameter distance $d$ and oriented to face the observer can be regarded as a straight string if the loop radius is greater than $d\Delta/2$ \cite{Vilenkin:1984ea}. In terms of loop length $L$ and comoving distance $\chi$, this means 
\begin{equation}
    L \, >\, 8\pi^2 G \mu \, a(t(\chi)) \, \chi,   \qquad \textrm{(minimum loop for simple lensing)}  \label{equation loop size criteria}
\end{equation}
where $a(t)$ is the scale factor at time $t$. This is valid in the vicinity of a loop, and here is where detectable lensing fringes are produced.

Therefore, the total fraction $\xi_l$ of the loop length that can be treated as a straight string (for a given $\chi$) is approximately the length distribution \Eq{equation loop distribution} integrated over the range that satisfies \Eq{equation loop size criteria} as
\begin{equation}
    \xi_{l}(\alpha, \Delta, \chi) \ \approx \ \frac{\int^{\alpha l(t)}_{8\pi^2 G \mu a(t) \chi} \, L/[L+\frac{1}{3} \Gamma G \mu l(t)]^2 \, dL}{\int^{\alpha l(t)}_{0} \, L/[L+\frac{1}{3} \Gamma G \mu l(t)]^2 \, dL}.
\end{equation}
Here, the upper bound comes from the longest loops at time $t(\chi)$. For the typical distance of 1 Gpc, the fraction is minimal $\xi_{l} \approx 0.97$ for $\alpha = 10^{-5}$ and $\Delta=10^{-6}$ and grows toward $\xi_l = 1$ for larger $\alpha$ and smaller $\Delta$. Thus, most of the total loop-length can be treated as a straight string. 

In all, in our final results, we assume $\xi_l = 1$ and use the energy fraction $\Omega_{\rm loop}/\Omega_{\rm cs}$ in \Eq{eq:OmegaRatio} with $\alpha=0.1$ \cite{Abbott:2017mem}.

\section{Lensing detection estimation} \label{section calculation}

\subsection{Detection criteria} \label{section criteria}

In this paper, we estimate the number of GW events with detectable lensing fringe in a simplified way. We measure the likelihood of the existence of fringes by a simple least chi-square test between the lensed and (best-fit) unlensed waveforms. It is based on the assumption that the un-fit differences can be well associated with the fringe because lensing fringes are characteristically different from other features of chirping GWs, as will be discussed in \Sec{sec:calculation}. This is our main assumption and simplification.

\medskip

First of all, the chirping GW itself must be detected confidently with large signal-to-noise ratio (SNR)
\beq
\textrm{SNR} \, \geq \,10.    \label{eq:crit1}
\eeq
Although lensed GWs will not be perfectly matched by unlensed waveforms in the discovery stage with matched filtering method, the overall waveform is still dominated by the chirping itself while lensing effects are subdominant perturbations. Thus, we assume that the discovery of lensed GWs using unlensed waveforms can still be well done. For simplicity, we use unlensed SNR.

Second, for confidently detected GWs, the log-likelihood $- \ln {\cal L}$ of the existence of lensing fringes is simply measured by the least chi-square method (see Appendix~\ref{app:likelihood} and \cite{Jung:2017flg,Dai:2018enj}) and required to be
\begin{equation} 
-2 \ln {\cal L} \,\equiv \,  4\int^{f_{\rm max}}_{f_{\rm min}} \frac{|\widetilde{h}^L(f)-\widetilde{h}_{\rm best-fit} (f)|^2}{S_n(f)} \ df \, > \, 9.     \label{eq:crit2}
\end{equation}
The $\widetilde{h}_{\rm best-fit} (f)$ is the unlensed GW waveform which minimizes the likelihood (see \Sec{sec:calculation} for unlensed waveforms and fitting parameters that we use), and $\widetilde{h}^L$ is the lensed waveform in \Eq{equation htilde lensed}; $S_n(f)$ is the power spectrum of the noise and $f_{\rm min,max}$ are the frequency range of the measurement. The large value of the $-\ln {\cal L}$ means that there are notable features in the GW that cannot be fit well by unlensed waveforms. Such features are assumed to be well associated with the fringe. Thus, $-\ln {\cal L}$ is equivalent to the likelihood of the fringe. 

Lastly, the fringe oscillation must be well resolved in the frequency domain
\beq
f_{\rm width} \, \geq \, 2 f_{\rm resol}.
\label{eq:criteria_width} 
\eeq
Otherwise, oscillations will be averaged out in measurements. Here, $f_{\rm width}$ is the width given by \Eq{equation fringe width analytic}, and $f_{\rm resol}$ is the frequency resolution of the measurement estimated by the discrete Fourier transform, hence given by the inverse of the total duration $T$ of a chirping measurement as
\begin{equation}
    f_{\rm resol}(f_{\rm max}, f_{\rm min}, \Mch_z) \= \frac{1}{T(f_{\rm max}, f_{\rm min}, \Mch_z)} \, \simeq \, 1.5 \, {\rm Hz}\, \times \left(\frac{\Mch_z}{100 \,M_\odot}\right)^{5/3} \left(\frac{f_{\rm min}}{10 \, \text{Hz}}\right)^{8/3}, \label{equation T}
\end{equation}
where $f_{\rm min} \ll f_{\rm max}$ is used in the last approximation and $\Mch_z$ is the redshifted chirp mass. The factor 2 in \Eq{eq:criteria_width} is a conservative factor that may account for additional uncertainties; we will later briefly discuss how our results change with this factor.

\medskip

With these simplified criteria in \Eq{eq:crit1} -- (\ref{eq:criteria_width}), the detection rate is calculated by integrating over all possible source and string properties (locations, masses, and etc) for detectable lensing. See Appendix~\ref{app:calc} for more details. The rate is a function of comoving merger-rate density, comoving cosmic string density $n_{cs}$, and detector specs $\equiv \{f_{\rm max}, f_{\rm min}, S_n(f)\}$. As for the merger-rate density, we assume all binary mergers are composed of two identical black holes (BHs) with the total mass $M$ (hereafter masses are always redshifted ones). Such comoving merger-rate density was calculated in \cite{Belczynski:2016obo} with the latest version in \cite{http://www.syntheticuniverse.org/stvsgwo.html}; we take the M10 optimistic prediction at $z_S = 0.1$, denoted as $n_s(M)$. $n_s(M)$ is given for several selected mass bins. The comoving cosmic string density is chosen to be 1 in a Hubble volume, $n_{cs} = (4.3\times 10^3 \ \text{Mpc})^{-3}$, by referring to the scaling solution~\cite{Caldwell:1991jj, Consiglio:2011qe,Harari:1987ht,Sikivie:1990sb}; if the string number density is greater than 1 in a Hubble volume, the detection rate will increase proportional to the number density. For simplicity, those densities are assumed to be isotropic and constant in redshift (although $n_s(M)$ at high redshift may be larger~\cite{Belczynski:2016obo}).

\subsection{Leading-order waveforms} \label{sec:calculation}

Following~\cite{Jung:2017flg}, we use the leading-order quadrupole chirping waveform in the inspiral stage. The observed waveform can be written in the form
\begin{equation}
    \widetilde{h}(f; A_0,\phi_0) \= A_0 f^{-7/6} e^{i \Psi (f,t_c=0) \+ i \phi_0},
    \label{equation htilde}
\end{equation}
for the purpose of calculating the likelihood in \Eq{eq:crit2}.
The frequency dependences of the chirping are explicitly shown in the amplitude and the phase
\beq
\Psi(f) \= 2\pi f t_c \+\frac{3}{128}(\pi {\cal M}_z f)^{-5/3},
\eeq
where $t_c=0$ (coalescence time) as discussed in regard of \Eq{eq:tc}. 
In the leading-order inspiral waveform, there are no other frequency-dependent features (see below for ignored effects). Thus, it is a good approximation to estimate the fringe detectability by minimizing $-\ln {\cal L}$ over two fitting parameters -- the overall amplitude $A_0$ and the overall phase shift $\phi_0$~\cite{Jung:2017flg,Dai:2018enj}. The inspiral-stage measurements are integrated up to the innermost stable circular orbit $f_{\rm max} = f_{\rm ISCO} = (3\sqrt{3}\pi M)^{-1}$ with the total binary mass $M$. 

However, the waveform in \Eq{equation htilde} ignores several effects that can produce oscillations in the waveform. First, the time-variations of detector orientations are ignored, either because they are almost constant during short measurement time or their time variations (daily or hourly) are very different from fringe oscillations. Second, the BH spin precession and orbital eccentricity are not considered. The former generally becomes quicker with inspiral (hence growing with frequency), and the latter quickly becomes small as the orbit circularizes during the chirping. Thus in general, they cannot mimic lensing fringes which have constant frequency widths and particular amplitudes. Last, the waveform dismisses the merger and ringdown phases as well as higher-order post-Newtonian effects. These become most relevant for relatively heavy binaries with total mass $M \gtrsim 50 \Msun$~\cite{Ajith:2008,Khan:2016}, while the majority of binary populations that we consider is lighter \cite{Belczynski:2016obo}. Thus, it is a good approximation to use leading-order waveforms with the least chi-square test for estimating the prospects of GW fringe detection. 

This simplified analysis was already used in Ref.~\cite{Jung:2017flg} for the point-mass lensing, and was found to be in reasonable agreement with more dedicated later works that even included BH spin effects~\cite{Christian:2018vsi,Dai:2018enj, Lai:2018rto}. Such good agreement stems again from the characteristic behaviors of fringes -- constant widths and particular amplitudes. Thus, we assume that the same simplifications can work well for the cosmic-string lensing too, while encouraging more dedicated analyses.

\section{Prospects at future detectors} \label{section prospects}

\begin{figure}[t]
    \centering
    \includegraphics[width=0.6\textwidth]{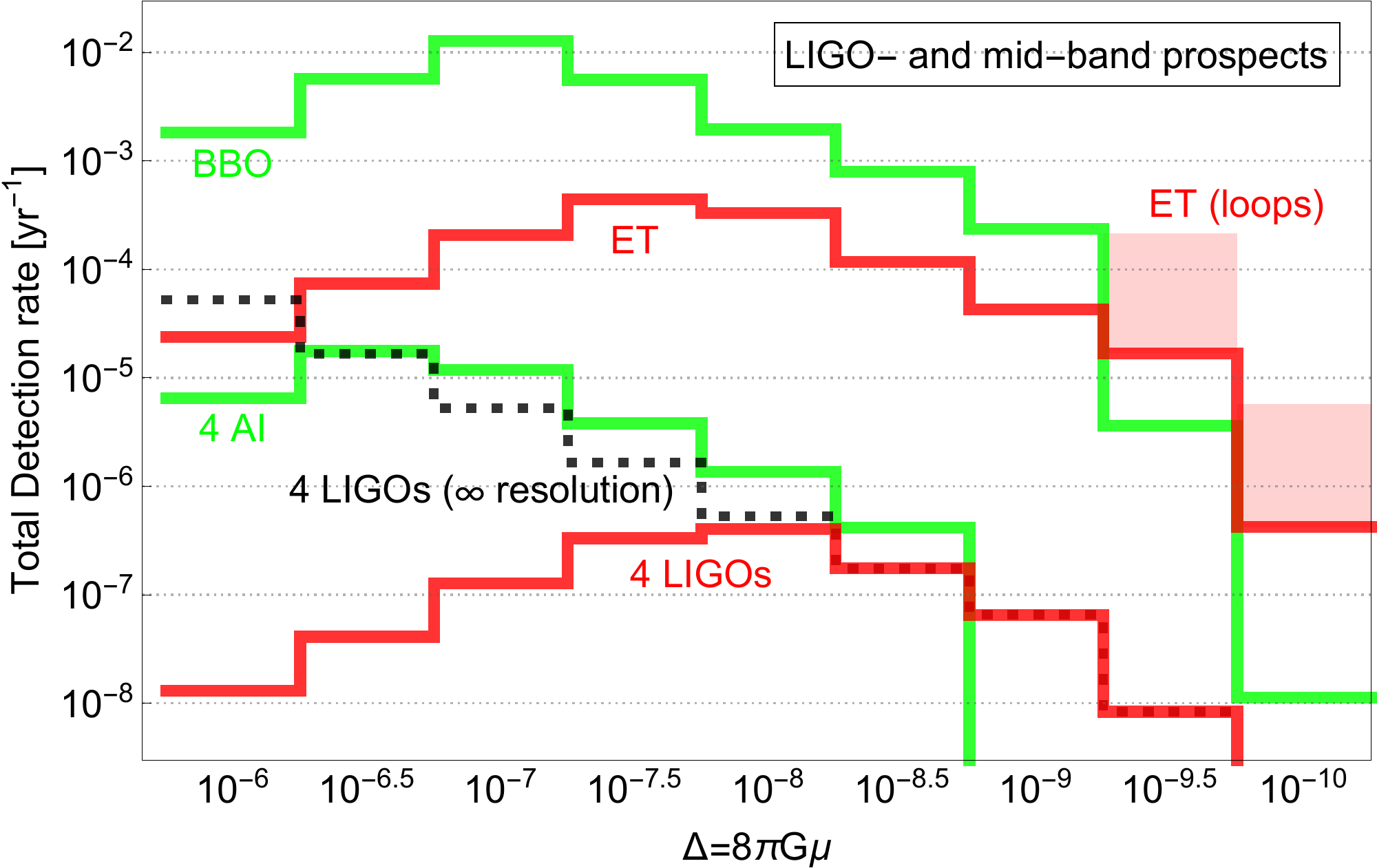}
    \caption{Expected detection rates of the GW fringe from cosmic strings at aLIGO, ET, AI, and BBO. The former two are high-frequency LIGO-band detectors (red), while the latter two are mid-band detectors (green). Last one-hour measurements in the mid-band detectors are used, while full measurements are used for the high-frequency detections. 4 detectors for aLIGO and AI are added, while 1 detector for ET and BBO. The black dashed is the aLIGO result with infinite frequency resolution. The red shaded region indicates additional loop contribution (if they exist) for ET. The cosmic string number abundance is 1 per horizon, and the GW merger rate is from \cite{Belczynski:2016obo,http://www.syntheticuniverse.org/stvsgwo.html}. The results are based on simplified analysis in \Sec{section calculation}.}
    \label{figure totrate}
\end{figure}

The estimated detection rates are shown in \Fig{figure totrate}. Shown benchmark detectors are high-frequency LIGO-band detectors (probing roughly 5 or 10 -- 5000 Hz) and mid-band detectors (roughly 0.1 -- 10 Hz). The former is represented by four advanced LIGO(aLIGO) detectors with design sensitivity~\cite{TheLIGOScientific:2016agk} (lower red) and one Einstein Telescope(ET) detector~\cite{Hild:2010id} (upper red), but ET has $10-50$ times smaller noise; the latter is by four Atom Interferometer(AI) detectors~\cite{Graham:2016plp,Graham:2017pmn} (lower green) and one Big Bang Observatory(BBO) detector~\cite{Yagi:2011wg} (upper green), but BBO has 10 -- 50 times smaller noise. Only last one-hour measurements of each GW in the mid-band are used (usually starting around 1 Hz) although GWs typically spend longer time in the mid-band, while full LIGO-band measurements (much shorter than an hour) are used. Only straight strings are included, but additional loop contributions that exist in local $U(1)$ models are shown for ET results (red shaded).

Above all, the overall detection rates of straight-string lensing fringes are much smaller than 1 per year. Thus, it is difficult to put constraints on the cosmic strings with just chirping GW observations. 4 aLIGOs are expected to detect $\sim 10^{-7} \, \text{yr}^{-1}$, 4 AIs $\sim 10^{-6} \, \text{yr}^{-1}$, ET $\sim 10^{-4} \, \text{yr}^{-1}$, and BBO $\sim 10^{-3} \, \text{yr}^{-1}$. The overall detection rates can be understood as follow. Since the majority of detectable lensing occurs for $|y| \lesssim \bigO(1)$, the lensing-detectable volume is $V \sim \Delta \cdot \chi_{\rm max}^3$  (where $\chi_{\rm max}$ is the comoving horizon distance of a detector; see \App{app:calc} for details), and consequently 
\beq
\text{detection rate} \, \sim \, (\Delta \cdot \chi_{\rm max}^3 n_{\rm cs})\times(n_{\rm s} \chi_{\rm max}^3).  \label{eq:quickrate}
\eeq
The first parenthesis, rewritten as $\Delta \cdot \chi_{\rm max}^3 / H_0^3 \,\sim {\cal O}(\Delta)$, is approximately the fraction of the Hubble volume that yields the detectable lensing, and the second is approximately the total GW merger rate. Using the GW merger rate density $300\text{/yr/Gpc}^3$~\cite{Belczynski:2016obo, http://www.syntheticuniverse.org/stvsgwo.html} and  $\chi_{\rm max} \= 1$ -- $3$ Gpc ($5$ -- $7$ Gpc) for aLIGO (ET)~\cite{http://www.et-gw.eu/index.php/etdsdocument}, we obtain total detection rate to be about $10^{-6}$ ($10^{-4}$) for $\Delta = 10^{-8}$, consistent with \Fig{figure totrate}.  Thus, the small $\Delta \ll 1$ which is already constrained by existing observables (hence, the small relic abundance) is one main reason for the small detection rate. But later we will discuss cases for increased detection rates, and that even a single event may be able to identify the cosmic string. 

For local $U(1)$ models, there are loops which can also produce GW fringes. Loop productions are already constrained for $\Delta \gtrsim 10^{-9.5}$ by stochastic GW searches~\cite{Abbott:2017mem}, so only results for $\Delta \leq 10^{-9.5}$ are presented. The results for ET (red shaded region in \Fig{figure totrate}) shows that loops can increase detection rates by an order of magnitudes, roughly by $\Omega_{\rm loop}/\Omega_{\rm cs}$ in \Eq{eq:OmegaRatio}. As discussed in \Sec{section loops}, this ratio is independent on the type of detectors, and $\alpha = 0.1$ and $\xi_l =1$ are used with logarithmic dependence on $\alpha$.

We emphasize that, unlike usual stochastic GW searches, the GW fringe is sensitive to straight strings regardless of the existence of loops. Thus, GW fringes can probe local $U(1)$ models as well as models without loops such as global $U(1)$ models, small-loop production cases, or other models with non-gravitational decays of strings. This property is complementary to the strong-lensing observables of light; they are compared in \Sec{sec:stronglight}. 

Our estimation is subject to uncertainties in the merger rate and string number density. For instance, the optimistic merger-rate density that we use (M10 model) is about 100-150 times larger than the M23 pessimistic merger-rate density~\cite{Belczynski:2016obo, http://www.syntheticuniverse.org/stvsgwo.html}. The cosmic string number density is also subject to ${\cal O}(10)$ variations among the simulations of string networks~\cite{Consiglio:2011qe}. 

\medskip

\Fig{figure totrate} also shows important complementary advantages of LIGO-band and mid-band detections. The large-$\Delta$ region is much better probed by mid-band detectors while the small-$\Delta$ region by LIGO-band detectors. The latter is because the fringe from small $\Delta$ is too broad to be detected by low-frequency GWs. The number of fringe oscillations in the frequency range up to $\sim f$ is given by ($y=1$ from \Eq{equation fringe width analytic})
\begin{equation}
    \text{Fringe number} \= \frac{fw}{\pi} \, \simeq \, \left( \frac{f}{\rm 200 \, Hz} \right) \left(\frac{d}{\rm 100 \, Mpc} \right) \left(\frac{\Delta}{10^{-9}} \right)^2, \label{equation fringe number}
\end{equation}
where all distances are of the same order $\sim d$. Thus, there is effectively no fringe in the mid-band for $\Delta \lesssim 10^{-9}$, while LIGO-band can probe down to $\Delta \sim 10^{-10}$.
\footnote{The region with Fringe number $< \mathcal{O}(1)$ is where the lensing fringe can be mimicked by various other oscillation effects discussed in \Sec{sec:calculation}, so the curves in \Fig{figure totrate} can be somewhat inaccurate in this region.}

In the large-$\Delta$ region, on the other hand, fringes become too narrow to be detected by short measurements. As discussed in regard of \Eq{eq:criteria_width}, the frequency resolution is estimated by the inverse of the total duration $T$ of a measurement. Thus, mid-band detections with $T=1$ hour (used in the figure) have better resolution than LIGO-band detections with $T \sim {\cal O}(1-10)$ seconds. \Fig{figure totrate} supports this explanation by showing the aLIGO results with artificially assumed infinite frequency resolution (black-dotted). The detection rate in this case indeed grows indefinitely with $\Delta$, showing that the frequency resolution is a limiting factor in the large-$\Delta$ region. If the resolution in the real analysis turns out to be different from the one that we use, it is the peak position of the detection rate that shifts accordingly, while the small-$\Delta$ results remain unchanged.

The complementary advantages give motivation to carry out combined broadband measurements which can achieve good sensitivities to both large and small $\Delta$ regions. Most full measurements in the broadband will be much longer than an hour that we used, so using longer measurements can also increase the overall SNR, and hence the detection rate. For example, one-week measurement with ET and BBO combined will give detection rate $\mathcal{O}(0.1\text{ -- }1) \, \text{yr}^{-1}$ for $\Delta \sim 10^{-6}$ and $\mathcal{O}(10^{-4}) \, \text{yr}^{-1}$ for $\Delta \sim 10^{-10}$. Thus, developing mid-band detections is one of the key steps to improve detection prospects of cosmic strings. 

\medskip

What about LISA and Pulsar Timing Array(PTA)? They probe lower-frequency ranges, $f = 0.001 \text{ -- } 0.1 \ \text{Hz}$ by LISA and $10^{-9} \text{ -- } 10^{-6}$ Hz by PTA. The LISA range can probe GW fringes from $\Delta \gtrsim 10^{-7} \text{ -- } 10^{-8}$. But the merger-rate of supermassive black-hole binaries that can be probed at LISA is too small, ${\cal O}(1-10)$ per year~\cite{Klein:2015hvg}. Thus, the fringe detection rate $\sim \Delta \times {\cal O}(1-10)$ is also likely to be too small. The estimation may also depend on accurate waveform models, which may receive sizable corrections beyond the leading quadrupole approximation for such massive binaries as discussed in \Sec{section criteria}. This is well beyond the scope of this work. On the other hand, the PTA range is too low to measure any GW fringe oscillations from $\Delta \lesssim 10^{-6}$. Thus, we conclude that the LIGO- and mid-bands are just appropriate to probe cosmic strings of $\Delta \lesssim 10^{-6}$.

\section{Discussion} \label{section discussion}

\subsection{Distinction from point-mass fringes} \label{section distinction}

We discuss how to distinguish cosmic-string lensing fringes from point-mass fringes using interference patterns. 

For $y>1$, the amplitude of the cosmic-string fringe decreases as the GW frequency increases because the diffracted ray has its amplitude proportional to $1/\sqrt{f}$. But this is not the case for the point-mass lensing or the cosmic-string lensing with $y\leq 1$ (interference is dominated by two geometric rays). Therefore, if a detected chirping GW has interference fringes that diminish with increasing $f$, this indicates that the lens is a cosmic string (with $y>1$). 

Cosmic strings with $y\leq 1$ can still be distinguished from point-mass lenses. In the frequency domain, the peak frequencies of fringes for a point-mass lens \cite{Nakamura:1999ptps, Takahashi:2003ix} are always $\pi/2$ shifted with respect to those of cosmic-string fringes \cite{Yoo:2012dn}, for a given fringe width. The shift is originated from the saddle-point contribution in the point-mass lensing, which does not exist in the cosmic-string lensing. The shift can be measured by correlating the peak frequency with the fringe width. In the case of the cosmic string (without the shift), $f_{\rm peak} = n f_{\rm width}$, while in the point-mass case (with the shift) the relation becomes $f_{\rm peak} = (n+\frac{1}{4}) f_{\rm width}$, where $n=$(0, 1, 2, ...). Thus, even with a single detection, such a correlation can identify a cosmic string as long as the signal shows at least two peaks with good enough frequency resolution.

\subsection{Prospect comparison with lensing of light}  \label{sec:stronglight}

We compare the GW fringe with existing light lensing observables: femto-lensing and strong lensing. All these have a nice property that they can probe straight strings independent on the existence of loops, but GW fringe is somewhat complementary to light observables.

The GW fringe is a GW counterpart of the photon femtolensing~\cite{Gould:1992}, but they can probe a different parameter space of cosmic strings. Both observables are lensing interference fringes. The GW fringe is naturally observed through the chirping of GW, while the femtolensing requires a stable and reliable photon energy spectrum of a source. As the wavelengths of photons and GWs are so different, their sensitivity ranges of $\Delta$ are different too. Lensing fringes are most readily recognized when the frequency range of observation spans $1-10^\#$ number of fringe oscillations, where the exact number $\# \sim {\cal O}(1)$ depends on the instruments. This can be written as $fw \sim 1 \text{ -- } 10^\#$ from \Eq{equation fringe number},
which gives the most sensitive range of LIGO-band detection as $\Delta \sim 10^{-8}$, of photon femtolensing with fast radio bursts (FRBs) as $\Delta \sim 10^{-11}$, and of GRBs as $\Delta \sim 10^{-17}$ \footnote{Similarly, for the point-mass lensing~\cite{Jung:2017flg}, the typical time-delay $\Delta t_d \simeq 4 G M_L$ gives the condition $ f \Delta t_d \, \simeq \, 2 \times 10^{-3} \left( \frac{M_L}{\Msun} \right) \, = \, 1 \sim 10^\# $. Thus, the LIGO observation is most sensitive to $M_L \sim 100 \Msun$~\cite{Jung:2017flg} while the GRB to $M_L \sim 10^{-14}\Msun$~\cite{Barnacka:2012bm,Katz:2018zrn,Jung:2019fcs}.}. Thus, GW fringes at LIGO- and mid-band are appropriate to probe $\Delta \lesssim 10^{-6}$ just below the current constraint,  and here is also where the largest cosmic-string abundance $\Omega_{cs} \simeq \Delta /3$ yields the highest detection rate. 

Cosmic strings can also be probed by the strong lensing of light: lensed images resolved in time or angle. The time-delay between lensed images is given by the path difference $w$ in \Eq{equation w}, while the angular separation is $\sim \Delta$. The best timing residual currently available is ${\cal O}$(ms) while the typical duration of, e.g., short GRBs is ${\cal O}(0.1)$ sec and FRBs is ${\cal O}(1)$ ms; the best angular resolutions are 0.1 arcsecond from James Webb and Hubble Space Telescope and 2 milliarcsecond from VLT interferometer while the bursts with compact sources appear much smaller. Thus, for the sources at cosmological distance $\sim$ Gpc, the 10-100 ms-timing accuracy can probe cosmic strings with $\Delta \gtrsim 10^{-10} - 10^{-9}$, and the milliarcsecond ($\sim 10^{-8}$ rad) resolution can probe $\Delta \gtrsim 10^{-8}$. The GW fringe is not only an independent probe, but its sensitivity range $\Delta \gtrsim 10^{-10}$ is also somewhat complementary in the smallest-$\Delta$ region, which in this case is limited by the highest frequency of the LIGO band. The detection rates of both light lensing and GW fringe are $\sim \Delta \times$(number of sources) up to some detection-related factors. So far, $\sim 100$ short GRBs, FRBs and $\sim 10-100$ GWs have been observed, but much more events will be observed in the near future. With $10^6 \text{ -- } 10^{10}$ future observations of cosmological bursts and/or binary mergers, one may be able to constrain $\Delta \,=\, 10^{-6} \text{ -- } 10^{-10}$, independent on the existence of loops.

\subsection{Robustness against astrophysical uncertainties} \label{sec:robust}

The GW fringe can be subject to various uncertainties from string movements, source movements, and source size which all can blur the sharp fringe pattern. Indeed, these are one of the main uncertainties of photon femtolensing. But GW fringe is much more robust against them, mainly due to much longer wavelengths and relatively short measurement time. 

The fringe pattern can be erased if the source size or the source/string movements (during measurement time) $\delta \ell \simeq d \Delta \delta y$  becomes large enough so that the change $\delta y \gtrsim 1$ sweeps a whole fringe period.
This limits the maximal source size or the source/string movements to be
\beq
\delta \ell \, \lesssim \,  \left( \frac{d}{\rm 100 \, Mpc} \right) \left( \frac{\Delta}{ 10^{-9}} \right) \cdot 3 \times 10^{15} {\rm km}.
\eeq
The limit $\sim 10^{15}$ km for the GW fringe is likely to be well satisfied for any sources and strings during even week-long measurements. But the limit becomes much stronger $\sim 10^6$ km for GRB femtolensing (with $\Delta \sim 10^{-18}$)~\cite{Yoo:2012dn,Matsunaga:2006uc}, and this may critically limit the robust observation of femtolensing. If cosmic strings can move relativistically, the size limit $\sim 10^6$ km implies the measurement time to be less than about a few minutes or so. 

The relativistic movement of strings can also induce the GW frequency red/blue-shift, but it is negligibly small of the order $\delta f/f \simeq \Delta$~\cite{Sazhina:2008}. It can also enhance the lensing deflection angle by the boost $\gamma = 1/\sqrt{1-v^2} \sim 1.3$~\cite{Shlaer:2005gk} (in average over the momentum direction) for a typical velocity $v\sim 2/3$ of relativistic strings. This effectively shifts the relevant $\Delta$ value by the $\gamma$ factor. Although we ignore this effect, one can account for this by using the shifted $\Delta$ in our results. In any case, a fringe pattern may change slightly but will not be erased.

Moreover, the GW waveform from a binary inspiral is well predicted by general relativity, governed most importantly by the binary masses. This allows detecting tiny GW fringes as well as testing general relativity~\cite{TheLIGOScientific:2016src} and probing weak (DM-induced) fifth forces~\cite{Choi:2018axi}. But astrophysical properties of GRBs or FRBs are under relatively poor control both theoretically and experimentally. 
Thus, the GW fringe is potentially a powerful precision observable of massive structures in the Universe.

\section{Conclusion}

We have studied the GW lensing fringe as a new probe of cosmic strings. We have shown that this interference fringe, observed in high-frequency LIGO-band and mid-band detectors, can be sensitive to a wide range of cosmic-string tension $\Delta = 8 \pi G\mu = 10^{-6} - 10^{-10}$ just below the current constraint from CMB anisotropies. In particular, mid-band detections yielding better frequency resolutions than those of the LIGO-band are required to probe large values of $\Delta$, while high-frequency LIGO-band detections are more suitable to small $\Delta$.
 
However, the detection rate of each detector is generally small, less than 1 per year, as it is proportional to $\Delta$ which is already constrained to be  $\lesssim 10^{-6}$. The small value of $\Delta$ implies the small abundance of the cosmic string $\Omega_{cs} \simeq \Delta /3$. But there are several cases where detection rates can be enhanced. Mid-band measurements which are typically much longer than an hour can enhance the estimated sensitivity to large $\Delta$, and the combination of LIGO- and mid-band measurements can further utilize synergies between them. If cosmic strings are produced from local $U(1)$ models, loops can produce an order-of-magnitude more GW fringe detections.

The GW fringe is complementary to existing observables. Unlike the well-known stochastic GW from loop decays, the GW fringe can directly probe straight strings that exist in any cosmic-string models. This allows to probe a more varieties of models. The strong lensing of lights shares this advantage with similar sensitivity ranges of $\Delta$ but with independent systematic uncertainties. Thus, the combination of GW fringes and light lensing observations, e.g. of short GRBs and FRBs, can probe $\Delta \lesssim 10^{-6}$ significantly better.

In any case, even a small number of fringes can be used for precision measurements of cosmic strings. Its fringes can not only be distinguished from those of other types of lenses, but also contain various information of strings. Our study has shown that \emph{long-time highest-frequency} measurements in the broadband $f \simeq 0.1-1000$ Hz are perhaps most appropriate for such precision capabilities~\cite{Graham:2017lmg,Nair:2018bxj,Isoyama:2018rjb,Takahashi:2003wm,Choi:2018axi}. The physics cases are ubiquitous too, including the early-Universe probe with cosmic strings as just one example, but also various DM candidates~\cite{Jung:2017flg,Nakamura:1997sw,Lai:2018rto} and larger-scale structures~\cite{Dai:2018enj}. It is always good to measure a wide range of frequency bands, but it would be more wonderful if one can utilize synergies achievable only through broadband measurements.

\begin{acknowledgments}

We thank Yanou Cui, David E. Morrissey, Takahiro Tanaka, Yi Wang and Chul-Moon Yoo for useful comments. Our work is supported by National Research Foundation of Korea under grant 2015R1A4A1042542, 2017R1D1A1B03030820, 2019R1C1C1010050, by Research Settlement Fund for the new faculty of Seoul National University, and SJ also by POSCO Science Fellowship. 

\end{acknowledgments}

\appendix

\section{More on $-\ln {\cal L}$ and relative likelihood}
\label{app:likelihood}

In this appendix, we explain that $-\ln {\cal L}$ in \Eq{equation htilde} corresponds to a relative likelihood. The discussion is based on Ref.~\cite{Jung:2017flg}, thanks to the anonymous referee; and similar discussion was also briefly made in Ref.~\cite{Dai:2018enj}.

A relative log-likelihood can be defined, for example, as
\beq
-2 \ln \Delta {\cal L}  \, \equiv \, \langle \widetilde{h}^L \,|\, \widetilde{h}^L \rangle \- \frac{ \langle \widetilde{h}^L \,|\, \widetilde{h}_{\rm best-fit} \rangle^2 }{ \langle \widetilde{h}_{\rm best-fit} \,|\, \widetilde{h}_{\rm best-fit} \rangle^2 },
\eeq
where $\langle \widetilde{h}_1 \,|\, \widetilde{h}_2 \rangle \,\equiv \, 4\, {\rm Re} \int df \, \widetilde{h}_1(f) \widetilde{h}_2^*(f) / S_n(f)$. This measures the possible improvement of matched-filtering significance one can obtain by using the correct lensed waveform compared to the best-fit unlensed waveform. One can conclude that if this improvement is large, unlensed waveform cannot match the lensed data well so that lensing can be detected. 

The relative likelihood $\Delta {\cal L}$ is equivalent to the ${\cal L}$ which can be rewritten in this notation as
\beq
-2 \ln {\cal L} \= \langle \widetilde{h}^L-  \widetilde{h}_{\rm best-fit}  \, | \, \widetilde{h}^L-  \widetilde{h}_{\rm best-fit} \rangle.
\eeq
First of all, \cite{Jung:2017flg} has numerically checked that $- \ln \Delta {\cal L}$ reproduces the same results obtained with $-\ln {\cal L}$. One can think of $\langle \widetilde{h}|$ as a vector in the two-dimensional space and $\langle h_1 | h_2 \rangle$ as an inner product. The angle between $\langle \widetilde{h}^L|$ and $\langle \widetilde{h}_{\rm best-fit}|$ is minimized by a best-fit analysis, but suppose the angle cannot be zero because lensed and unlensed waveforms cannot match perfectly. For the given angle, the relative likelihood yields the component of $\langle \widetilde{h}^L|$ perpendicular to the direction of $\langle \widetilde{h}_{\rm best-fit}|$. The length of this component, which is minimized by a best-fit, is the relative likelihood of lensed data and unlensed best-fit waveform. On the other hand, $- \ln {\cal L}$ gives the same answer in a slightly different way. For the same given angle (which is thought to be found by a best-fit), the length of the difference vector $\langle \widetilde{h}^L-  \widetilde{h}_{\rm best-fit}|$ is minimized when this vector is perpendicular to the direction of $\langle \widetilde{h}_{\rm best-fit}|$. Thus, a proper best-fit minimization analysis will yield the same results with both definitions.

\section{Detection rate calculation} \label{app:calc}

In this section, we outline a procedure of calculating the lensing detection rate. All distances and volumes in this section are comoving ones.

\begin{figure}[t]
    \centering
    \includegraphics[width=0.48\textwidth]{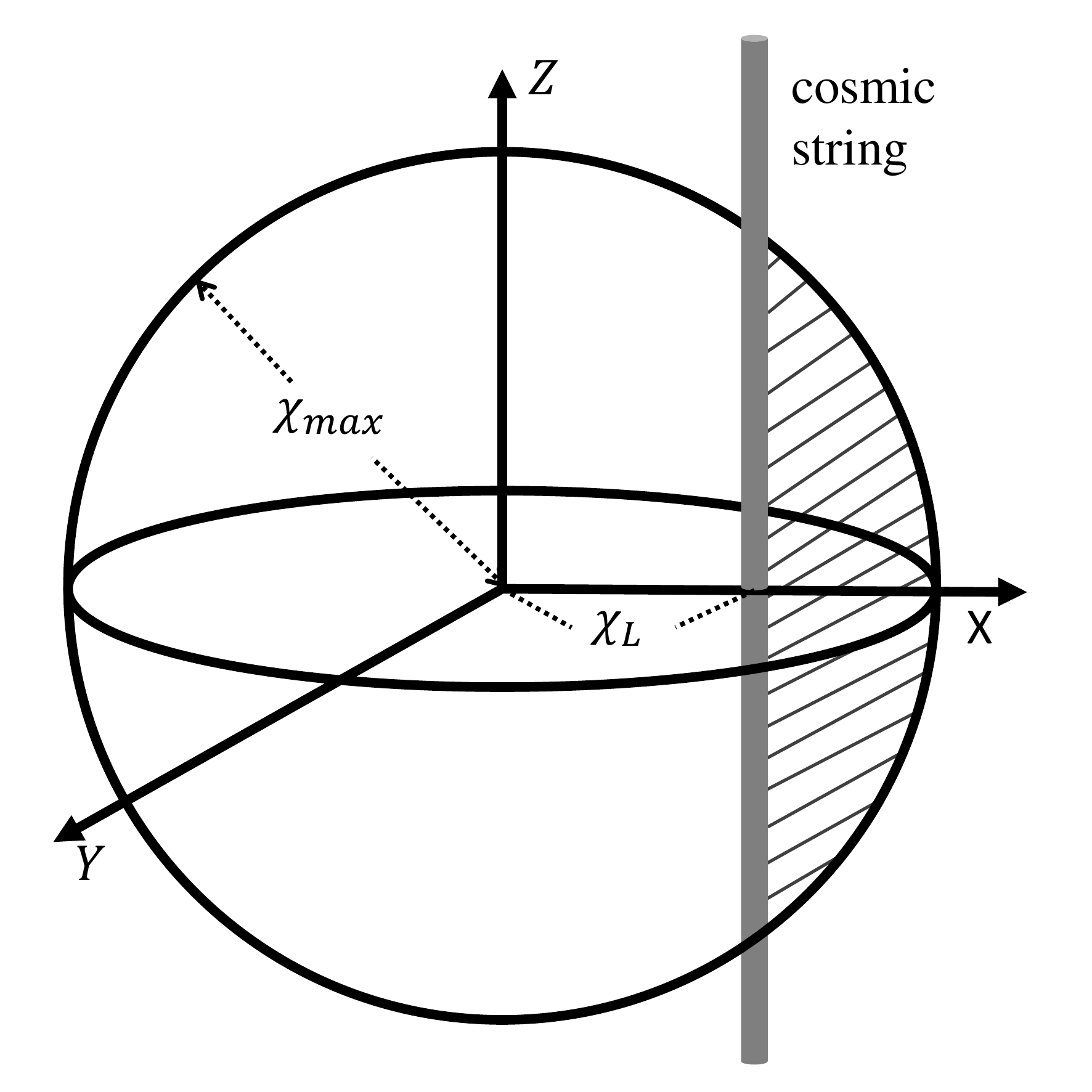}
    \caption{The comoving coordinate system and the placement of a cosmic string used for lensing detection calculation. The hatched region indicates the possible source positions that can give detectable lensing.} 
    \label{figure coord. sys.}
\end{figure}

The detection rates in \Fig{figure totrate} and \Eq{eq:quickrate} are calculated by
\begin{equation}
    \text{detection rate}(\Delta, M; f_{\rm max}, f_{\rm min}, S_n(f)) \= n_s(M) \, n_{cs} \, V_{6D}(\Delta, M; f_{\rm max}, f_{\rm min}, S_n(f)), \label{equation event rate}
\end{equation}
where $V_{6D}$ is the 6-dimensional(6D) volume of the locations of source and string for detectable lensing (satisfying the three criteria in \Sec{section criteria}). This is expressed as a product between comoving densities and the volume, as densities are isotropic and homogeneous.
Because of the isotropy, the $V_{6D}$ is further given by a volume integral of the 3D volume $V_s$ of the source position that produces detectable fringes for the string at $\chi_L$
\begin{equation}
V_{6D}(\Delta) \= \int^{\infty}_{0} 4\pi \chi_L^2 \, V_s (\chi_L, \Delta) \, d\chi_L. \label{equation V6D}
\end{equation}
Here, the string direction does not matter due to the isotropy, and its distance is defined to be the closest one as shown in \Fig{figure coord. sys.}.

The $V_s$ is the volume of the hatched region in \Fig{figure coord. sys.}. The maximum source distance $\chi_{\rm max}$ is determined by the observability of the GW, i.e. SNR $>10$ condition. If a string is put on the $XZ$ plane as in the figure, for each $(X,Z)$ location, there is a maximum $Y_{\rm max} (X,Z)$ that can satisfy detection criteria. Then 
\begin{equation}
V_s(\chi_L, \Delta) \= 2 \iint^{\chi_{\rm max}} Y_{\rm max}(X,Z) \, dX dZ, \label{equation V}
\end{equation}
where the factor 2 accounts for $Y<0$ and the integration is over the hatched region. $Y_{\rm max}$ is determined by the remaining detection criteria (chi-square test and the frequency resolution). 

Just for more technical details, from \Eq{equation F} and \Eq{equation fringe width analytic}, the two parameters that essentially determine the criteria are $w$ and $y$. So instead of evaluating the criteria in the 3D space of $(X,Y,Z)$, it is easier to integrate over the 2D plane of $(w, y)$. This gives $y_{\rm max}$ as a function of $w$. Thus, for a given $(X, Z)$, we first compute $w$ (note that $Y_{\rm max} \ll X \text{ or } Z$) and then convert $y_{\rm max} (w)$ to $Y_{\rm max} (X, Z)$ by using $y = 2\tan^{-1}\left[Y/(X-\chi_L)\right]/\Delta$. 
In numerical integration, $X$ and $Z$ axes are divided by 16 sectors with rectangular quadrature, while the Boolean condition $\sqrt{X^2+Z^2}\leq \chi_{\rm max}$ is imposed. Similarly, $V_{6D}$ integral is done over rectangular quadratures with 16 sectors for $0 \leq \chi_{L} \leq \chi_{\rm max}$.

\section{Detection rate for each mass bins}
\label{app:RateVsM}

\begin{figure}[t]
  \centering
  \includegraphics[width=0.48\textwidth]{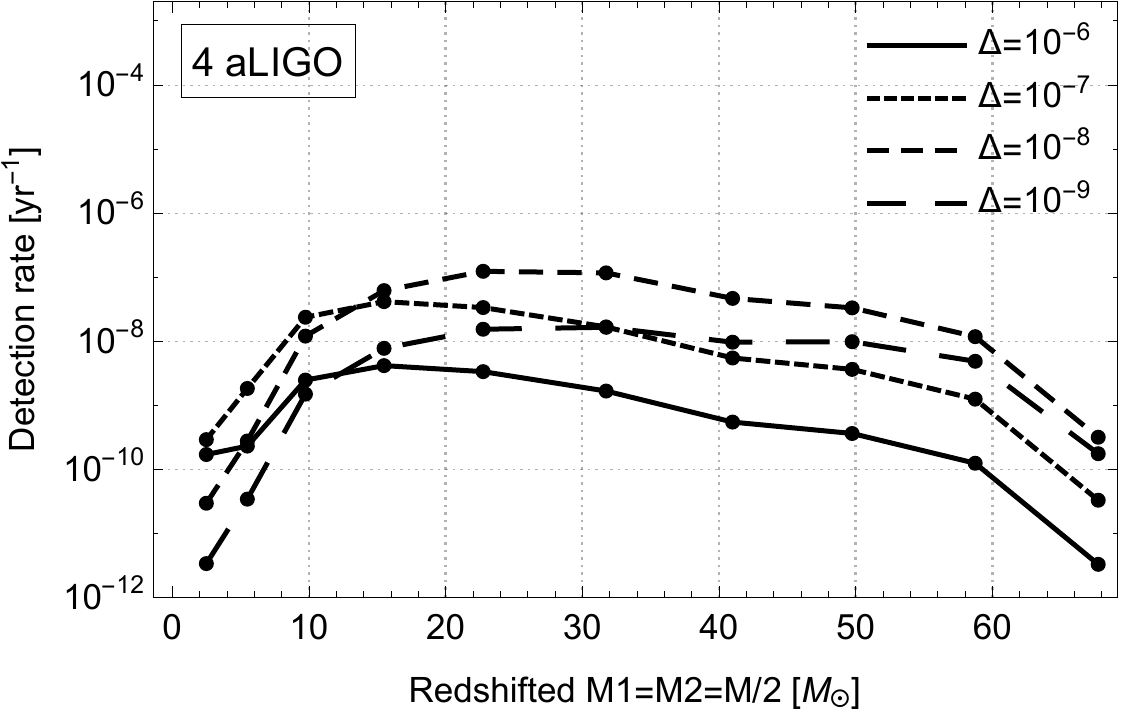}
  \hfill
  \includegraphics[width=0.48\textwidth]{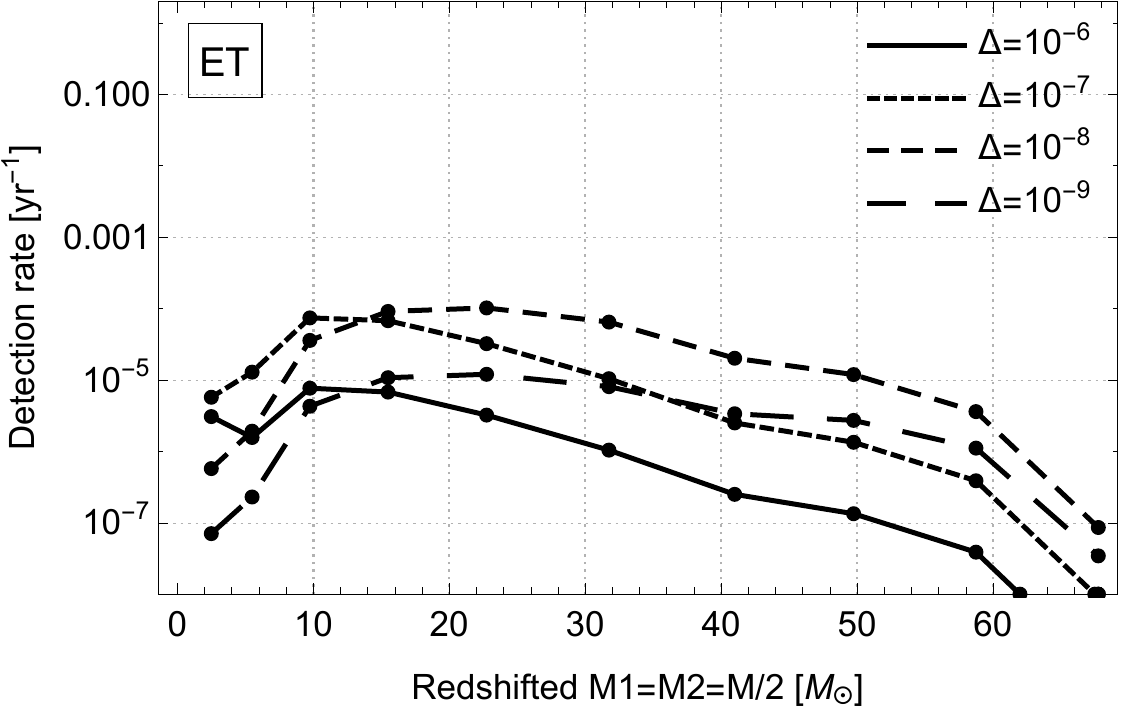}
  \caption{
  Detection rate of the GW fringe with four aLIGOs (left) and one ET (right) in each binary mass bin. Several values of string tensions $\Delta=8\pi G \mu = 10^{-6,-7,-8,-9}$ are shown.}
  \label{figure MvsRate}
\end{figure}

In \Fig{figure MvsRate}, we show LIGO-band detection rate of each binary mass bin. First of all, the detection rates typically decrease in both small and large masses. This is because small-mass binaries produce weak GWs, whereas large-mass binaries have low merger-rate densities~\cite{Belczynski:2016obo, http://www.syntheticuniverse.org/stvsgwo.html}. Secondly, the overall detection rates grow with $\Delta$, except for large $\Delta \simeq 10^{-7}$ -- $10^{-6}$. This is in accordance with \Sec{section prospects}, but the suppression is more severe for heavy binaries, which spend relatively shorter time in the LIGO-band.


\end{document}